\documentclass[aps, prl, superscriptaddress, twocolumn, amsfonts, amsmath, amssymb, floatfix]{revtex4-1}

\usepackage{graphicx}
\usepackage{dcolumn}
\usepackage{bm}
\usepackage{epsfig}
\usepackage{braket}
\usepackage{color}
\usepackage{ulem}
\usepackage{amsmath}
\usepackage[colorlinks=true, letterpaper=true, pdfstartview=FitV, linkcolor=blue, citecolor=blue, urlcolor=blue]{hyperref}

\begin{document}

\title{Spin squeezing and many-body dipolar dynamics in optical lattice clocks}

\author{Chunlei Qu}
\author{Ana M. Rey}
\affiliation{JILA, NIST, and Department of Physics, University of Colorado, Boulder, Colorado 80309, U.S.A.}
\affiliation{Center for Theory of Quantum Matter, University of Colorado, Boulder, Colorado 80309, U.S.A.}

\date{\today}

\begin{abstract}
The recent experimental realization of a three-dimensional (3D) optical lattice clock not only reduces the influence of collisional interactions on the clock's accuracy but also provides a promising platform for studying dipolar many-body quantum physics. Here, by solving the governing master equation, we investigate the role of both elastic and dissipative long-range interactions in the clock's dynamics and study its  dependence on lattice spacing, dimensionality, and dipolar orientation. For small lattice spacing, i.e., $k_0a\ll 1$, where $a$ is the lattice constant and $k_0$ is the transition wavenumber, a sizable spin squeezing appears in the transient state which is favored in a head-to-tail dipolar configuration in 1D systems and a side-by-side configuration in 2D systems, respectively. For large lattice spacing, i.e., $k_0a\gg 1$, the single atomic decay rate can be effectively suppressed due to the destructive dissipative emission of neighboring atoms in both 1D and 2D. Our results will not only aid in the design of the future generation of ultraprecise atomic clocks but also illuminates the rich many-body physics exhibited by radiating dipolar system.
\end{abstract}
	
\maketitle

Alkaline-earth-metal atoms have recently attracted an intensive research interest in the cold-atom community~\cite{Ludlow2015,Daley2011,Cazalilla2014} as they can be used for the development of atomic clocks with unprecedented stability and accuracy~\cite{Daley2011,Hinkley2013,Bloom2014}. Typical one-dimensional (1D) optical lattice clocks suffer from systematic frequency shifts induced by the atomic collisions~\cite{Daley2011, Rey2014, Ludlow2011} and this has stimulated the built up of next-generation optical lattice clocks in a deep 3D lattice loaded with at most one atom per site~\cite{Campbell2017}. In this regime, the clock becomes immune to atomic collisions, however, atoms can still interact via long-range dipolar interactions which yield non-negligible frequency shifts and thus can impact the performance of the clock~\cite{Chang2004}. Understanding these interactions is thus not only fundamental to avoid undesirable systematic frequency shifts to improve the clock performance~\cite{Chang2004,Kramer2016} but also important for us to take advantage of them for the generation of entangled states, such as spin-squeezed states, which are useful for enhanced metrology.

\begin{figure}[b!]
\centerline{
\includegraphics[width=8.6cm]{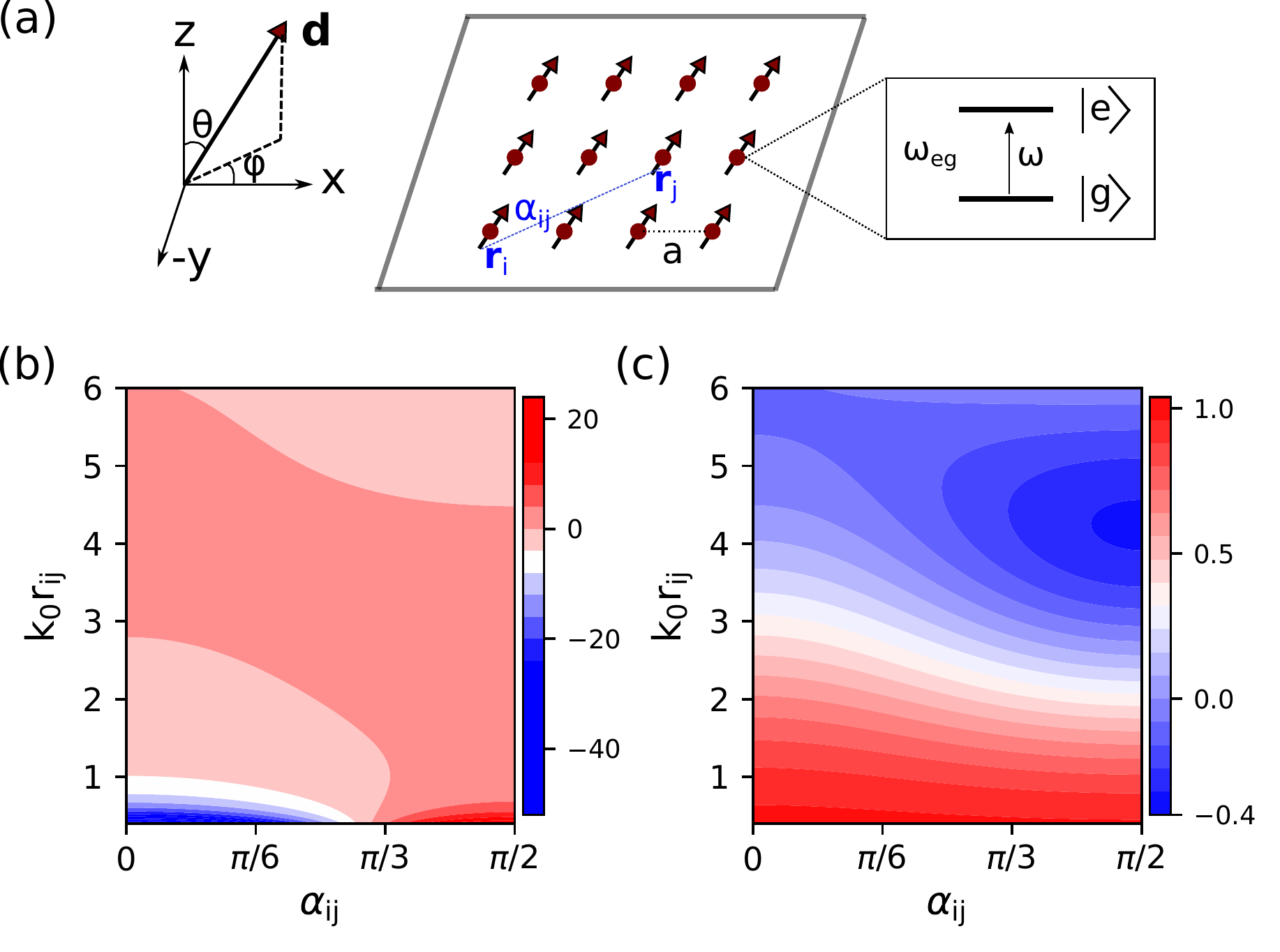}}
\caption{(a) Illustration of the optical lattice clock where $a$ is the lattice constant and $\alpha_{ij}$ is the angle between the dipole polarization direction $\mathbf{d}=d(\sin\theta\cos\varphi, \sin\theta\sin\varphi, \cos\theta)$ and the position vector $\mathbf{r}_{ij}$ connecting the two atoms at sites $i$ and $j$. Each atom can be approximated as a two-level system with the ground state and excited state labeled by $|g\rangle$ and $|e\rangle$. [(b) and (c) Profile of the scaled elastic interaction $g(\mathbf{r}_{ij})/\Gamma$ and dissipative interaction $f(\mathbf{r}_{ij})/\Gamma$ as a function of the angle $\alpha_{ij}$ and $k_0r_{ij}$, where $k_0$ is the wave number of the clock interrogation laser and $r_{ij}$ is the distance between the two atoms. It is worth noting that $f(\mathbf{r}_{ij})/\Gamma\rightarrow 1$ as $k_0r_{ij}\to 0$, reaching a collective emission limit.}
\label{Fig:fg}
\end{figure}

Spin-squeezed states are a class of quantum states having suppressed spin variance along a certain direction, at the cost of enhanced variance along the orthogonal directions~\cite{Wineland1992,Kitagawa1993,Ma2011}. These states offer a pathway  to overcome the standard quantum limit and to improve the accuracy of phase measurements~\cite{Pezze2018}. Various quantum systems, such as trapped ions~\cite{Sackett2000,Leibfried2004,Leibfried2005,Monz2011, Bohnet2016}, Bose-Einstein condensates~\cite{Esteve2008, Gross2010, Riedel2010}, and cold thermal atoms~\cite{Appel2009, Leroux2010a, SchleierSmith2010, Cox2016, Hosten2016}, have been explored to generate spin-squeezed states. In this work, we study spin-squeezing in the optical lattice clocks and focus on the competition between dipolar interaction and dissipation. Previous work has shown that, without dissipation, the long-range dipole-dipole interaction can generate spin-squeezing in the transient state~\cite{Schachenmayer2015,FossFeig2016}. Furthermore, in the absence of elastic dipolar interaction, collective spontaneous emission may be engineered to yield a spin-squeezed steady state by an external coherent driving~\cite{Tudela2013}. Here, we show that a sizable spin-squeezing can be achieved in the transient states  without an external drive in the small lattice  spacing regime where the elastic dipolar interaction is much stronger than the dissipation. With increasing lattice  spacing, the elastic dipolar interaction becomes comparable to the dissipation and the spin-squeezing disappears. In this regime, however, we find that the decay rate of the atomic array can be effectively suppressed compared to their single atom radiative decay rate due to destructive dissipative emission of nearest neighbors (a type of subradiance effect~\cite{Gross1982}). The change of the decay rate is approximately given by $Mf_{12}$, where $M$ is the number of nearest neighbors and $f_{12}$ is the cooperative emission rate between these neighbors.

We consider $N$ two-level atoms which can be treated as point dipoles pinned in an optical lattice with only one atom per site. After eliminating the electromagnetic field modes, one obtains the master equation governing the dynamics of the atomic degree of freedom~\cite{Lehmberg1970,Gross1982,Olmos2013,Ostermann2013}
\begin{equation}
\frac{d\rho}{dt} = - \frac{i}{\hbar} [H_0, \rho] + \mathcal{L}_f[\rho], 
\end{equation}
where the two terms determine the coherent evolution and dissipation of the system, respectively. The Hamiltonian that governs the coherent dynamics reads
\begin{equation}
H_0 = -\frac{\hbar}{2}\sum_i (\Omega e^{i\mathbf{k}_R\cdot \mathbf{r}_i}\sigma_i^+ + H.c.)  + \frac{\hbar}{2}\sum_{i\neq j} g(\mathbf{r}_{ij}) \sigma_i^+ \sigma_j^-
\end{equation}
The first term  describes the laser drive, resonant with the atomic transition and with Rabi frequency $\Omega$ and laser wavevector $k_R$. In the following we will assume this term  drives the atomic transition at the beginning of the Ramsey protocol and then is turned off during the dark time dynamics when only dipolar interactions are present. The long-range elastic dipolar interaction, which exchanges excitations between two atoms at $\mathbf{r}_i$ and $\mathbf{r}_j$, is characterized by $g(\mathbf{r}_{ij})$ with $\mathbf{r}_{ij}=\mathbf{r}_i-\mathbf{r}_j$. The spin raising operator for the atom at site $i$ is $\sigma_i^+=|e_i\rangle \langle g_i|$ where $|g_i\rangle$($|e_i\rangle$) is the ground (excited) state of the atom. The dissipative interaction which describes the processes of both independent and cooperative decay is described by the Lindblad operator which is of the following form
\begin{equation}
\mathcal{L}_f[\rho] = \frac{1}{2}\sum_{i,j}f(\mathbf{r}_{ij})(2\sigma_j^- \rho \sigma_i^+ - \sigma_i^+ \sigma_j^-\rho - \rho\sigma_i^+\sigma_j^-).
\end{equation}

The coefficients of the elastic and dissipative dipolar interactions have the following explicit forms
\begin{widetext}
\begin{eqnarray}
g(\mathbf{r}_{ij}) &=& -\frac{3\Gamma}{2} \left\{
\sin^2\alpha_{ij} \frac{\cos\zeta_{ij}}{\zeta_{ij}} + (3\cos^2\alpha_{ij}-1)\left[
\frac{\cos\zeta_{ij}}{\zeta_{ij}^3} + \frac{\sin\zeta_{ij}}{\zeta_{ij}^2}
\right]
\right\} \label{eq:fg1}
\\
f(\mathbf{r}_{ij}) &=& \frac{3\Gamma}{2}\left\{
\sin^2\alpha_{ij}\frac{\sin\zeta_{ij}}{\zeta_{ij}} + (3\cos^2\alpha_{ij}-1)\left[
\frac{\sin\zeta_{ij}}{\zeta_{ij}^3} - \frac{\cos\zeta_{ij}}{\zeta_{ij}^2}
\right]
\right\}
\label{eq:fg2}
\end{eqnarray}
\end{widetext}
where $\Gamma=k_0^3d^2/3\pi\epsilon_0\hbar$ is the single atom spontaneous emission rate, $k_0$ is the transition wave number, $d=|\mathbf{d}|$ is the dipole matrix element between the ground and excited states, $\zeta_{ij}=k_0|\mathbf{r}_{ij}|=k_0a\sqrt{\sum_{\mu}(i_\mu-j_\mu)^2}$ ($\mu=x,y,z$) characterizes the dimensionless distance between two atoms, $a$ is the lattice constant, and $\alpha_{ij}$ is the corresponding angle between the dipole polarization direction $\mathbf{d}$  and the vector $\mathbf{r}_{ij}$ connecting the two atoms [see Fig.~\ref{Fig:fg}(a)].

The profile of the elastic and dissipative interaction is shown in Figs.~\ref{Fig:fg}(b) and \ref{Fig:fg}(c). When the distance between the two atoms is small compared with the transition wavelength, i.e., when $k_0r_{ij} \ll 1$, the short-range ($1/r^3$) term dominates in the elastic dipolar interaction. An exception occurs at the so-called magic angle $\alpha_m=\arccos(1/\sqrt{3})\approx 54.7^{\circ}$ where the ($1/r^3$) and ($1/r^2$) terms vanish and the elastic dipolar interaction becomes much smaller. The inelastic interaction, however, becomes homogeneous at small distance with $f(\mathbf{r}_{ij})\to\Gamma$, reaching the limit of collective superradiant emission. When the distance between the  atoms is large compared to the dipolar transition wavelength , i.e., when $k_0 r_{ij} > 1$, the magnitudes of the elastic and dissipative interactions are comparable and the sign of the dissipative interaction can become negative [blue area in Fig.~\ref{Fig:fg}(c)]. As we will show in the following, the negative nearest-neighbor dissipative interaction can effectively reduce the single atom decay rate.

The initial state of the system after a $\pi/2$ Rabi pulse is a coherent spin state with all the atoms prepared in an equal superposition of the ground and excited states
\begin{equation}
|\psi(t=0)\rangle = \otimes_{j=1}^N \frac{1}{\sqrt{2}}(|g_j\rangle + e^{i\phi_j}|e_j\rangle )
\label{eq:wf}
\end{equation}
where $\phi_j=\mathbf{k}_R\cdot\mathbf{r}_j$ is the phase accumulated along the propagation direction of the Rabi pulse. In the following, we set $\exp(i\phi_j)=1$. This can be achieved by either considering a low-dimensional system with the Rabi pulse propagating perpendicular to the atomic array/plane or by choosing a wave vector $k_R$ such that $\phi_j=2\pi n$ where $n$ is an integer. For the sake of simplicity, we shall focus on the following low-dimensional systems: (i) a 1D array along $x$ and (ii) a 2D lattice in the $x-y$ plane [see Fig.~\ref{Fig:fg}(a)]. Without loss of generality, we assume that the dipole polarization direction is in the $x-z$ plane ($\mathbf{d}=d\{\sin\theta,0,\cos\theta\})$ i.e., $\varphi=0$  as shown in  Fig.~\ref{Fig:fg}). Consequently, for two dipoles at sites $i$ and $j$ along the $y$ direction, we have $\alpha_{ij}=\pi/2$ and $\alpha_{ij}=\pi/2-\theta$ if they are along the $x$ direction.

\begin{figure}[t!]
\centerline{
\includegraphics[width=8.6cm]{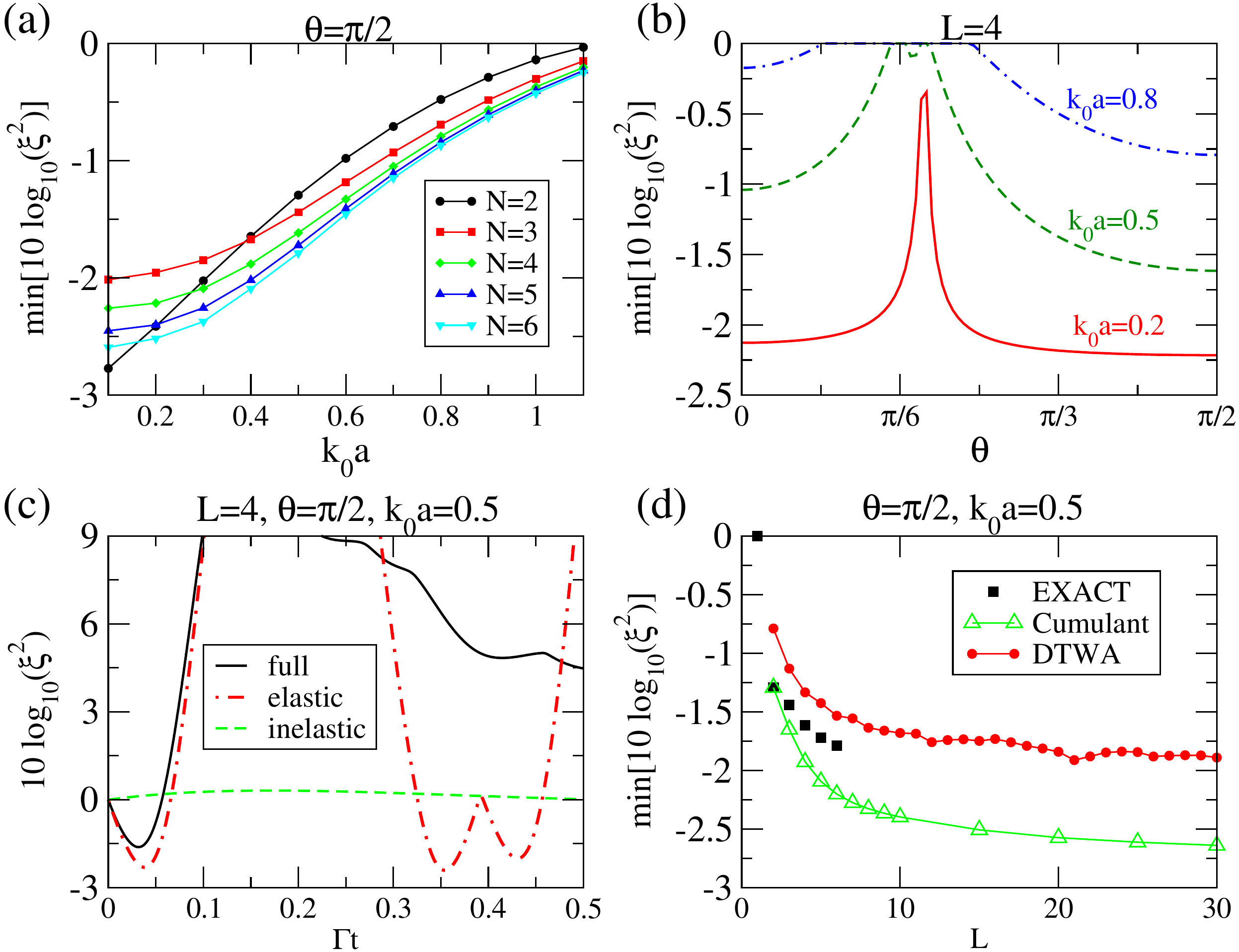}}
\caption{Spin squeezing obtained in the transient state of atoms in a 1D array. (a) Plot of the minimum squeezing as a function of atomic distance $k_0a$ for $N=2$ (circles), $3$ (squares), $4$ (diamonds), $5$ (up triangles), and $6$ (down triangles). (b) Plot of the minimum squeezing as a function of the dipole polarization orientation $\theta$ where the squeezing drastically disappears around the magic angle $\theta_m$. The system size is $N=4$ and the atomic spacing $k_0a=0.8$ (blue dash-dotted line), $0.5$ (green dashed line), and $0.2$ (red solid line). (c) Time dependence of the squeezing for an $L=4$ system with only elastic interaction (red dash-dotted line), only dissipative interaction (green dashed line), and with both interactions (black solid line) at present. (d) Dependence of the minimum squeezing on the system size $L$ of the 1D array obtained from the exact calculation (black squares), DTWA (red circles), and cumulant expansion (green triangles).}
\label{Fig:1D}
\end{figure}

Mathematically, the squeezing parameter of a spin system can be calculated as
\begin{equation}
\xi = \text{min}_{\mathbf{n}}\frac{\sqrt{N}\sqrt{\langle (\mathbf{S}\cdot\mathbf{n})^2 \rangle - \langle \mathbf{S}\cdot \mathbf{n} \rangle ^2}}{|\langle \mathbf{S}\rangle|}
\end{equation}
where the minimization is over unit vector $\mathbf{n}$ perpendicular to the mean spin direction $\langle \mathbf{S} \rangle = \sum_i\langle \mathbf{s}_i\rangle$ with $ s_i^\mu = \sigma_i^\mu /2$ ($\mu=X,Y,Z$) as the $\mu$-component of the spin operator for the $i$-th atom.

Let us first consider a 1D array system with $L$ sites (thus $N=L$ atoms). The dipolar many-body dynamics can be obtained by solving the master equation directly for a system with a few atoms (here up to $N=6$). This is relevant for state-of-the-art experiments working with optical tweezers where small arrays of alkaline earth atoms can be trapped~\cite{tweezer1,tweezer2}. Our exact numerical calculation indicates that the system quickly evolves into a spin-squeezed state and the squeezing disappears for longer evolution time. For $N=2$, the minimum squeezing obtained is smaller for smaller values of $k_0a$ as the dipolar interaction becomes stronger. For $N>2$, the dipolar interaction between non-nearest neighbors contribute to the squeezing in a nonuniform manner, causing a different behavior for the squeezing compared to that for $N=2$ [Fig.~\ref{Fig:1D}(a)]. Still, for $N>2$, the optimal squeezing improves  with increasing particle number  $N$ and smaller interparticle separation $k_0a$. The squeezing tends to disappear for larger values of $k_0a\sim 1$ where the dissipative interaction becomes comparable to the elastic interaction.

Since the dipolar interaction is sensitive to the dipole orientation $\theta$ (note that $\alpha_{ij}=\pi/2-\theta$ in 1D), we plot the minimum squeezing as a function of angle $\theta$ for different values of $k_0a$ in Fig.~\ref{Fig:1D}(b). Consistently  with the magnitude  of the elastic dipolar interactions set by  $g(\mathbf{r}_{ij})$ [Fig.~\ref{Fig:fg}(c)], the squeezing tends to disappear near the magic angle $\theta_m=\pi/2-\alpha_m\approx 35.3^\circ$ and the variation of the minimum squeezing as a function of angle $\theta$ becomes more pronounced for smaller $k_0a$. To better understand the role of the elastic and dissipative interaction on the spin squeezing, we show the time dependence of the spin squeezing for a system with $N=4$ atoms in Fig.~\ref{Fig:1D}(c). Spin squeezing is generated for all the three cases presented where the squeezing is as expected  optimal for a system with only elastic interactions and surprisingly tiny but not zero if there is only dissipative interaction. In fact, in the absence of elastic interactions, the optimal spin squeezing could be $\gtrapprox$1dB when the dissipative interaction becomes homogeneous (collective emission). For the real system, it is the interplay between both elastic and dissipative interactions that dictates the achievable squeezing.

Since the Hilbert space scales as $4^N$, an exact solution of the master equation becomes impractical for a larger system size and hence we seek an approximate numerical method, the recently developed discrete truncated Wigner approximation (DTWA)~\cite{Schachenmayer2015}. The equation of motion for an operator $A$ is given by $dA/dt=\text{Tr}[\dot{\rho}A]$ which usually involves multiple-point correlators. In DTWA, the closed set of differential equations are obtained within the mean-field approximation where only one-point correlators are involved~\cite{supp}. To take into account quantum fluctuations, the mean-field equations are evolved for many different initial conditions chosen by randomly sample the initial quantum spin distribution function. For the initial state considered in this work, this is done by fixing $\langle s_i^X\rangle =1/2$ and randomly choosing $\langle s_i^Y\rangle, \langle s_i^Z\rangle =\pm 1/2$. The expectation values are obtained by averaging the results of the corresponding observable over all the initial samplings.

For comparison, we also study the system with the cumulant expansion method~\cite{supp} where the closed set of equations involve both one-point and two-point correlators. The three-point correlators and higher-order correlators are approximately decomposed into one-point and two-point correlators~\cite{supp}. In a wide parameter regime, these two methods give qualitatively similar results. However, our simulation also indicates that the cumulant expansion method is not adequate to describe the system when entanglement becomes relevant, in our case manifested as significant spin squeezing. The variance of the average spin [the square root term in Eq. (7)] may become negative and thus the squeezing pathologically diverges. In this parameter regime, dissipation is less important and the DTWA can be trusted, at the least at the qualitative level as demonstrated in Ref.~\cite{Schachenmayer2015}.

In Fig.~\ref{Fig:1D}(d), we show the optimal squeezing obtained in the 1D system for different system sizes. Despite that there is a discrepancy between the three methods, DTWA still qualitativelly captures the behavior of the minimum squeezing for small atom number. The value of the minimum squeezing quickly saturates as a function of atom number $N$, indicating that, at $k_0a=0.5$, the interaction between two far-separated atoms is too weak to contribute to the squeezing. For much smaller values of $k_0a$, more atoms can be packed in the same spatial region and interactions between close-by atoms become stronger. Hence, a better optimal spin squeezing can be obtained. To obtain the spin squeezing more accurately, the DTWA method could be generalized by including the two-point and even higher-order quantum correlations in the equations of motion~\cite{Pucci2016,Orioli2017}.

\begin{figure}[t!]
\centerline{
\includegraphics[width=8.6cm]{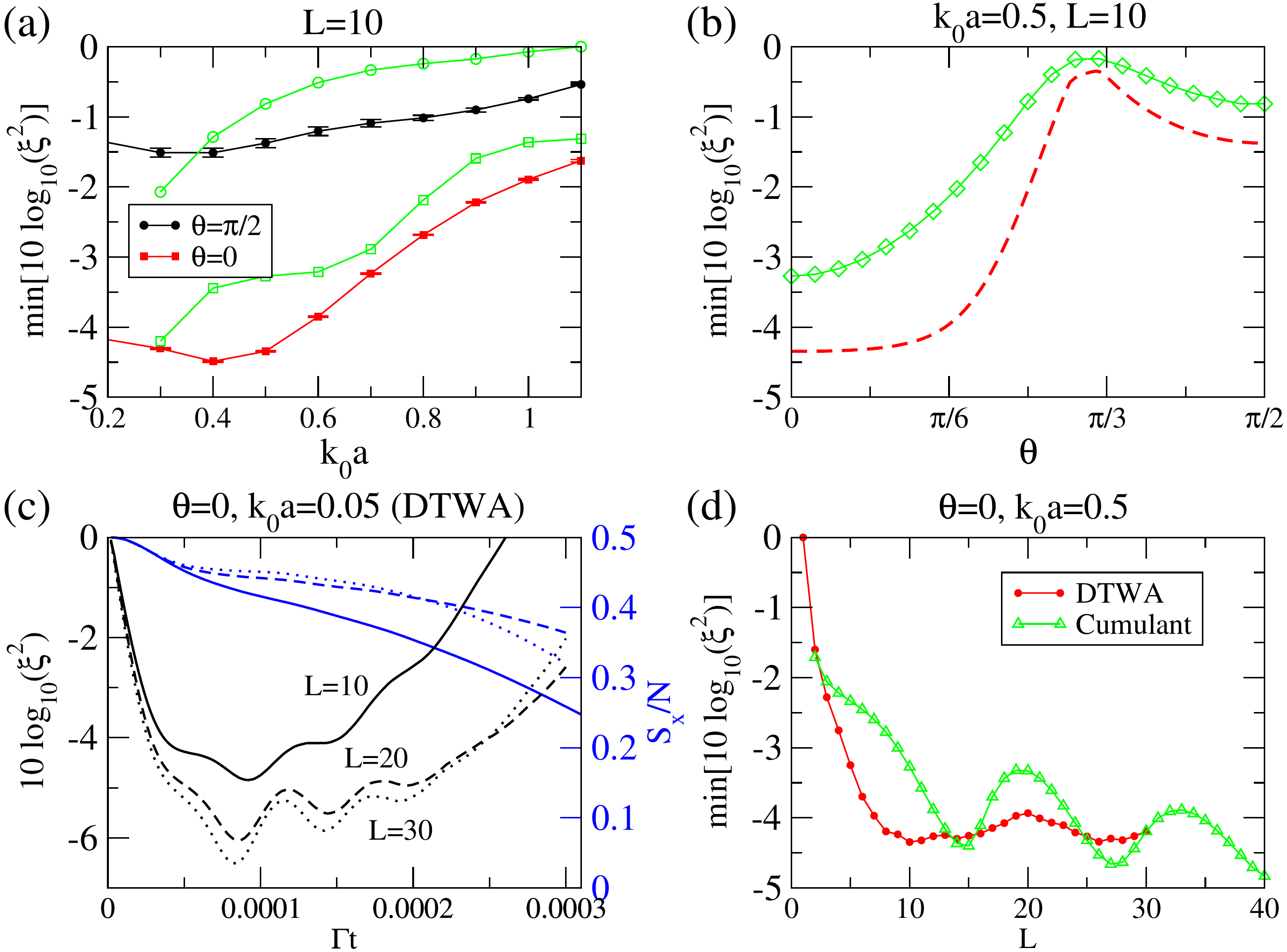}}
\caption{Spin squeezing in the transient state of the 2D optical lattice clock. The results are obtained with the DTWA method. (a) Plot of the minimum squeezing as a function of the atomic distance $k_0a$ for $\theta=0$ (red squares) and $\theta=\pi/2$ (black circles). (b) Plot of the minimum squeezing as a function of the dipole polarization orientation $\theta$ (note that we have taken $\varphi=0$). (c) Time dependence of the squeezing (left axis) and the average magnetization $\langle S^X\rangle/N$ (right axis) for a 2D system with $L=10$ (solid lines),$20$ (dashed lines), and $30$ (dotted lines) at small atomic spacing $k_0a=0.05$. (d) Dependence of the minimum squeezing on the system size $L$ of the 2D system. The error bars in panel (a) indicate the uncertainty of the DTWA results due to the finite number of sampling for the initial state. In panels (a), (b), and (d), the results obtained with cumulant expansion (green symbols) have been included for comparison.}
\label{Fig:2D}
\end{figure}

For a 2D square lattice with atom number $N=L^2$, where $L$ is the dimension of the system along one direction, the dipolar interaction becomes anisotropic, i.e., $\alpha_{ij}$ are usually different between different pairs of atoms. To investigate the spin squeezing of a large-sized 2D system, we adopt the DTWA method~\cite{supp}. As shown in Fig.~\ref{Fig:2D}(a), the minimum squeezing of a 2D system still becomes optimal at small values of $k_0a$. However, the dependence of squeezing on $k_0a$ is nonmonotonous and the optimal squeezing could be as large as $4.5$dB at $k_0a=0.5$ and $\theta=0$. The fact that the system exhibits a much better squeezing at $\theta=0$ is related to the fact that all the dipoles are aligned in a side-by-side configuration in the 2D plane, giving rise to a more isotropic dipolar interaction between any two nearest neighbors. At $\theta=\pi/2$, the orientation of the dipoles is side-by-side along $y$ and head-to-tail along $x$. Hence, the dipolar interaction between different neighbors may cancel each other due to its anisotropic character and spin squeezing is not favored.

As shown in Fig.~\ref{Fig:2D}(b), the dependence of the optimal  squeezing on the angle $\theta$ resembles the behavior of  the 1D system. However, the value of the angle $\theta$ for which the spin squeezing disappears is different from 1D. This is because the magic angles $\alpha_m$ cannot be reached simultaneously along all  directions in the 2D plane.

Recently, a spin-ordered phase has been identified for the 2D spin system in the presence of a $1/r^3$ dipole-dipole interaction~\cite{Kwasigroch2017}. In this phase, the observable quantity $\langle S^X\rangle/N$ first decays for a very short time and then remains at a certain value for a long time. In our system, at $k_0a\ll 1$, signature of a similar spin-ordered phase shows up even in  the presence of  collective  emission ($f_{ij}\approx\Gamma$). Our calculation shows that the spin-ordered phase is accompanied by a long lasting and oscillatory spin-squeezing which becomes prominent for larger systems [Fig.~\ref{Fig:2D}(c)]. In fact our numerical simulations in 2D suggest that squeezing survives longer for larger system sizes, an observation that could have important implications for the metrological usefulness of dense dipole arrays~\cite{Qu2}. Nevertheless, the dissipative interaction eventually kills this spin-ordered phase. Similarly to the dependence of the spin squeezing on the system size in 1D [Fig.~\ref{Fig:1D}(d)], the minimum spin squeezing also quickly saturates in 2D with increasing system size(Fig.~\ref{Fig:2D}(d)). In the cumulant expansion, the optimal spin squeezing also quickly saturates. However, it exhibits an oscillatory behavior which is likely due to the interferences caused by finite-size effects.

In the presence of strong  dissipation, the excited atoms inevitably decay to their ground state. For noninteracting atoms, the decay of each atom is solely determined by the spontaneous emission constant $\Gamma$ with $\langle s^X \rangle = \frac{1}{2}e^{-\Gamma t/2}$. Analytic solutions can also be obtained for an interacting two-atom system. The closed equations of motion are
\begin{eqnarray}
\frac{d}{dt}\langle \sigma_i^+ \rangle  &=& -\frac{\Gamma}{2}\Gamma \langle \sigma_i^+ \rangle + \frac{1}{2}(f_{ij}-ig_{ij})\langle \sigma_j^+\sigma_i^Z \rangle \\
\frac{d}{dt}\langle \sigma_i^+ \sigma_j^Z \rangle &=& -\frac{3\Gamma}{2}\langle \sigma_i^+ \sigma_j^Z \rangle - \Gamma \langle \sigma_i^+ \rangle - \frac{1}{2}(f_{ij}+ig_{ij})\langle \sigma_j^+ \rangle \nonumber \\
&& - f_{ij} \langle \sigma_j^+ \sigma_i^Z \rangle
\end{eqnarray}
To study the role of the dissipative interaction, it is instructive to consider the situation where the elastic interaction $g_{ij}=0$. Under this assumption, the exact solution has the following simple form
\begin{equation}
\langle S^X \rangle/N =\frac{1}{2}e^{-\frac{1}{2}(\Gamma+f_{ij})t}\left(
1-\frac{f_{ij}}{2\Gamma} e^{-\Gamma t} + \frac{f_{ij}}{2\Gamma}
\right)
\end{equation}
which decays exponentially at the rate of $\Gamma+f_{ij}$. Consequently, the decay rate becomes slower than the single atom spontaneous decay rate $\Gamma$ if the inelastic interaction is negative $f_{ij}<0$, under the assumption that the elastic interaction $g_{ij}\approx 0$. This condition can be fulfilled for large values of $k_0r_{ij}$, for example, when $\alpha_{ij}=\pi/2$ and $k_0r_{ij}\approx 4.23$ (see Fig.~\ref{Fig:fg}(c)). From the full expression of $g_{ij}$ and $f_{ij}$ [Eqs.~(\ref{eq:fg1}) and (\ref{eq:fg2})], the optimal value of $k_0r_{ij}$ that gives a vanishing elastic interaction and negative dissipation can be analytically estimated by letting $\sin{\zeta_{ij}}=-1$ which gives $k_0r_{ij}=\zeta_{ij}=3\pi/2\approx 4.7$, in agreement with the above exact numerical value. The slow decay dynamics is similar to but also different from the previously studied subradiance effect~\cite{subradiance1, subradiance2, subradiance3}. The familiar subradiance effect, occurs when the atoms are closely packed and are prepared in an antisymmetric entangled state. In our case, it occurs when the lattice constant is comparable to the wavelength of the interrogated transition and for an initial coherent spin states, both accessible in current clock experiments.

\begin{figure}[t!]
\centerline{
\includegraphics[width=8.6cm]{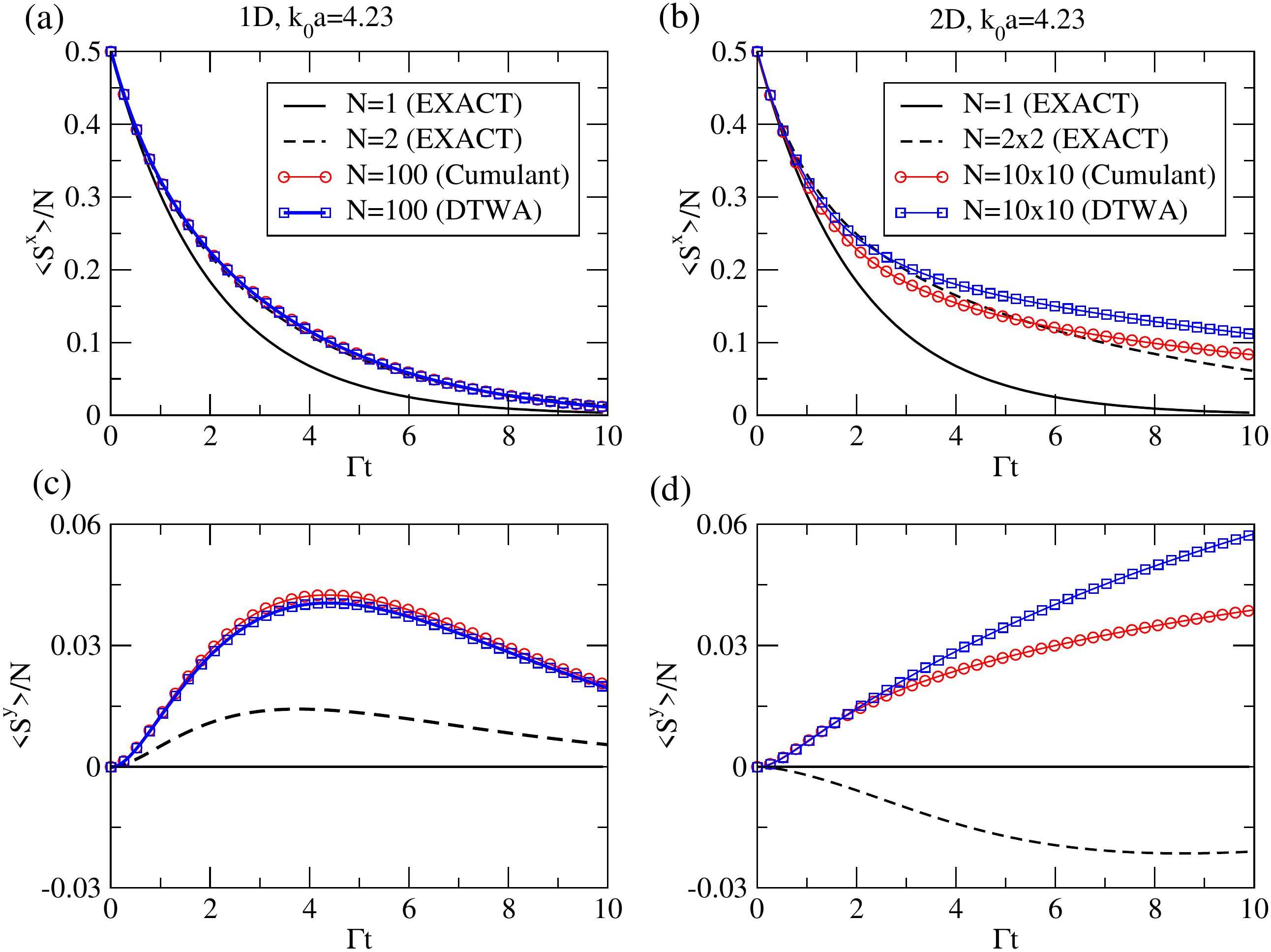}}
\caption{[(a) and (c)] Time dependence of $\langle S^X\rangle/N$ and $\langle S^Y\rangle/N$ for a 1D system with particle number $N=1$ (black solid lines), $N=2$ (black dashed lines) and $N=100$ (blue squares) at $k_0a=4.23$. [(b) and (d)] The same results for a 2D system.}
\label{Fig:slow1}
\end{figure}

For a system with many atoms, the decay rate depends on the total number of nearest-neighboring atoms and the geometry of the system. For example, as shown in Figs.~\ref{Fig:slow1}(a) and \ref{Fig:slow1}(b), the decay for an $N=4$ system in 2D is much slower than that of an $N=2$ system in 1D. For a larger system in 2D, all the neighbors contribute to the decay dynamics in a different way which complicates the dipolar dynamics. The multiple contributions give rise to different behaviors at short and long evolution times for an $N=10^2$ system in 2D (see Fig.~\ref{Fig:slow1}(b))~\cite{Henriet2019}.

\begin{figure}[t!]
\centerline{
\includegraphics[width=8.6cm]{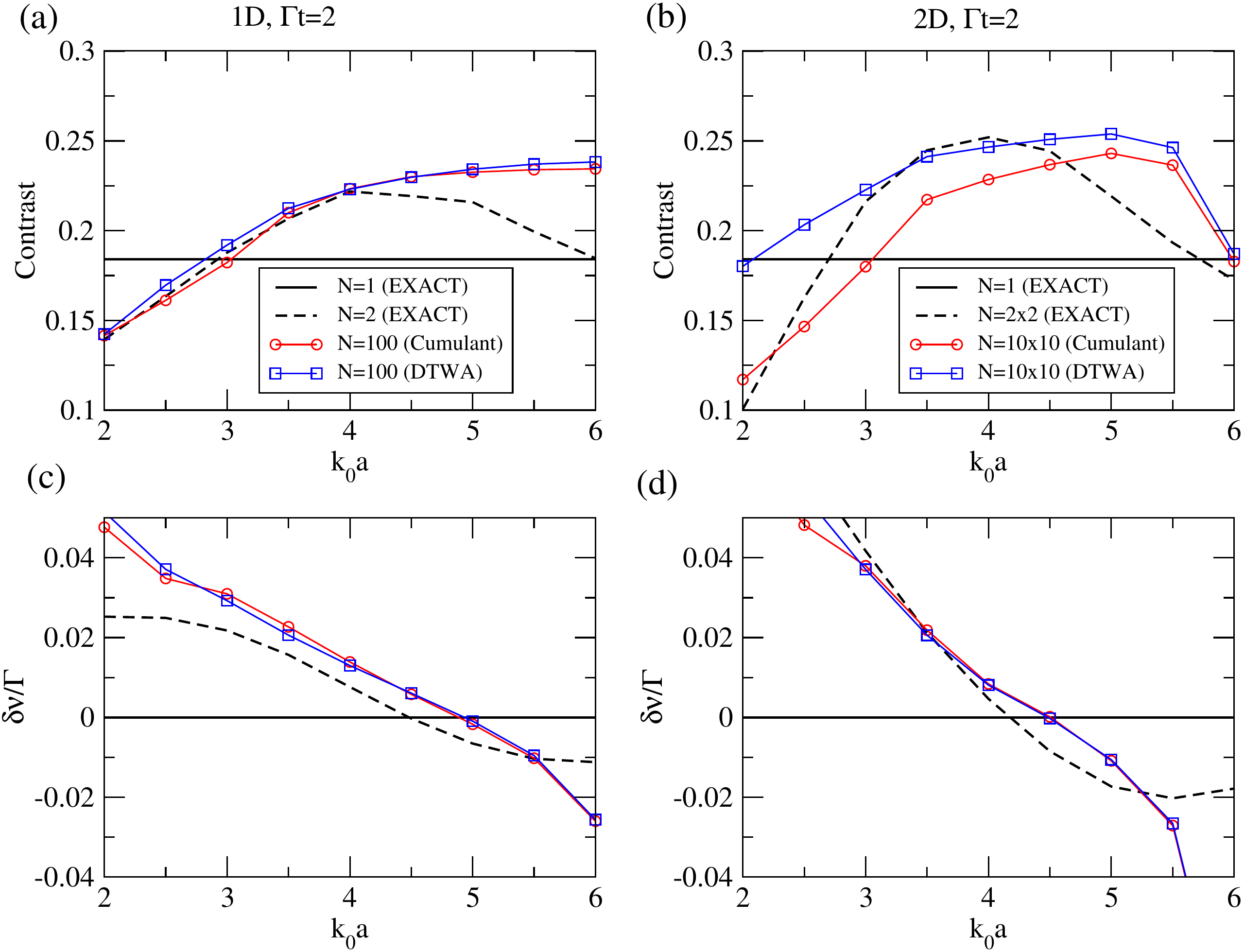}}
\caption{[(a) and (c)] Contrast and frequency shift as a function of $k_0a$ at $\Gamma t=2$ for a 1D system with particle number $N=1$ (black solid lines), $N=2$ (black dashed lines) and $N=100$ (red circles and blue squares). [(b) and (d)] The same results for a 2D system.}
\label{Fig:slow2}
\end{figure}

In optical lattice clock experiments~\cite{Bromley2018}, two observables are frequently measured: (i) the Ramsey fringe contrast, $\mathcal{C}=2\sqrt{\langle S^X \rangle ^2 + \langle S^Y \rangle ^2}/N$, and (ii) the density-dependent frequency shift $\delta\nu$ which can be calculated as $\tan(\delta\nu 2\pi t)=\langle S^Y \rangle / \langle S^X\rangle$. For a two-particle system,
the contrast and frequency shift can be obtained analytically~\cite{supp}. In Fig.~\ref{Fig:slow2}, we present the numerical results for the contrast and the frequency shift as a function of $k_0a$. The contrast of the few-atom systems ($N=2$ in 1D and $N=4$ in 2D) exhibits a peak around $k_0a=4.23$ where the elastic dipolar interaction is very weak and the corresponding frequency shift is very close to zero. This offers a parameter regime where the clock's performance can be improved. For systems with a larger atom number, the contrast and the frequency shift are qualitatively the same, demonstrating again that the interaction from the next nearest neighbors and beyond plays a small role in the dipolar dynamics.

To observe the phenomena explored in this work, one can consider the bosonic alkaline-earth atoms which have zero nuclear spin and thus simple atomic structure~\cite{Bromley2016}. For example, the $^1S_0\leftrightarrow {^3}P_1$ transition of $^{88}$Sr atoms has a transition wavelength $\lambda=689$nm and a natural linewidth $\Gamma=2\pi\times 7.5$kHz. The degeneracy of the three $J=1$ levels can be lifted by a magnetic field. For typical lattice constant $a\sim 400$nm, the constant $k_0a\approx 3.66$ and hence $f(k_0a)<0$. This setup provides a platform for the observation of the slower decay dynamics. To observe the large spin squeezing, one can instead choose the transition between the $^3P_0$ and $^3D_1$ states where the corresponding transition wavelength is $\lambda=2.6\mu$m and the decay rate $\Gamma=2\pi\times 290$kHz. At magic wavelength, the lattice is identical for the two states and the lattice constant $a=206.4$nm, the parameter $k_0a$ could be as small as $0.25$~\cite{Olmos2013}, reaching the regime to observe a large spin squeezing in a 2D system.

In conclusion, we have shown that the alkaline-earth-metal atoms in optical lattices provide a platform for studying dipolar many-body quantum physics. Particularly, we have identified two regimes (i.e., $k_0a\ll 1$ and $k_0a\gg 1$) where different many-body dynamics emerge and strongly depend on the lattice spacing and the dipolar orientation. To enhance the dipolar interactions the use of subwavelength lattices could be an interesting direction~\cite{Yi2008,Lacki2016,Jendrzejewski2016,Wang2018}. A generalization of the current work is to keep the external coherent Rabi driving on during the dipolar dynamics where one may expect richer dynamics and nontrivial steady states emerging from the cooperation and competition between the drive, the elastic, and the dissipative interactions.

\begin{acknowledgments}
We thank A. Kaufman, G. Juzeli\=unas, A. Pi\~{n}eiro Orioli, and A. Cidrim for helpful discussions. This work is supported by the Air Force Office of Scientific Research Grant No. FA9550-18-1-0319 and its Multidisciplinary University Research Initiative grant (MURI), by the Defense Advanced Research Projects Agency (DARPA) and Army Research Office Grants No. W911NF-16-1-0576 and No. W911NF-19-1-0210, the National Science Foundation Grant No. PHY-1820885, JILA-NSF grant PFC-1734006, and the National Institute of Standards and Technology.
\end{acknowledgments}

\end{document}